\documentclass[fleqn,usenatbib]{mnras}

\usepackage{newtxtext,newtxmath}

\usepackage[T1]{fontenc}

\DeclareRobustCommand{\VAN}[3]{#2}
\let\VANthebibliography\thebibliography
\def\thebibliography{\DeclareRobustCommand{\VAN}[3]{##3}\VANthebibliography}

\usepackage{graphicx}	
\usepackage{amsmath}	

\usepackage{amssymb}	
\usepackage{booktabs, caption, makecell}

\usepackage{threeparttable}
\usepackage{ulem}

\title[The first low-mass binary from TMTS]{The first low-mass eclipsing binary within the fully convective zone from TMTS}

\author[C. Liu et al.]{
Cheng Liu,$^{1}$\thanks{E-mail: liucheng@bjp.org.cn (CL)}
Xiaofeng Wang,$^{2,1}$\thanks{E-mail: wang\_xf@tsinghua.edu.cn (XW)}
Xiaobing Zhang,$^{3}$
Mikhail Kovalev,$^{4}$
Jie Lin,$^{2,5}$
Gaobo Xi,$^{2}$
Jun Mo,$^{2}$\newauthor
Gaici Li,$^{2}$
Haowei Peng,$^{2}$
Xin Li,$^{1}$
Qiqi Xia,$^{2}$
Abdusamatjan Iskandar,$^{6}$
Xiangyun Zeng,$^{7}$
Letian Wang,$^{6}$\newauthor
Liying Zhu,$^{4}$
Xuan Song,$^{1}$
Jincheng Guo,$^{1}$
Xiaojun Jiang,$^{3}$
Shengyu Yan,$^{2}$
Jicheng Zhang$^{8}$
\\
$^{1}$Department of Scientific Research, Beijing Planetarium, Beijing 100044, China\\
$^{2}$Physics Department and Tsinghua Center for Astrophysics, Tsinghua University, Beijing 100084, China\\
$^{3}$CAS Key Laboratory of Optical Astronomy, National Astronomical Observatories, Chinese Academy of Sciences, Beijing 100101, China\\
$^{4}$Yunnan Observatories, China Academy of Sciences, 650216 Kunming, China\\
$^{5}$School of Astronomy and Space Sciences, University of Science and Technology of China, Hefei, 230026, China\\
$^{6}$Xinjiang Astronomical Observatory, Chinese Academy of Sciences, Urumqi, Xinjiang 830011, People's Republic of China \\
$^{7}$Center for Astronomy and Space Sciences, China Three Gorges University, Yichang 443000, China \\
$^{8}$Department of Astronomy, Beijing Normal University, Beijing, 100875, China
}

\date{Accepted 2024 May 3; Received 2024 April 24; in original form 2023 October 31}

\pubyear{2024}

\begin{document}
\label{firstpage}
\pagerange{\pageref{firstpage}--\pageref{lastpage}}
\maketitle

\begin{abstract}
We present a comprehensive photometric and spectroscopic analysis of the short-period ($\sim$5.32 hours) and low-mass eclipsing binary TMTSJ0803 discovered by Tsinghua-Ma Huateng Telescope for Survey (TMTS). By fitting the light curves and radial velocity data with the Wilson--Devinney code, we find that the binary is composed of two late spotted active M dwarfs below the fully convective boundary. This is supported by the discovery of a significant Balmer emission lines in the LAMOST spectrum and prominent coronal X-ray emission. In comparison with the typical luminosity of rapidly rotating fully convective stars, the much brighter X-ray luminosity ($L_{X}/L_{\rm{bol}} = 0.0159 \pm 0.0059$) suggests the stellar magnetic activity of fully convective stars could be enhanced in such a close binary system. Given the metallicity of [M/H] = $-$ 0.35 dex as inferred from the LAMOST spectrum, we measure the masses and radii of both stars to be $M_{1} = 0.169 \pm 0.010~M_{\odot}$, $M_{2} = 0.162 \pm 0.016~M_{\odot}$, $R_{1} = 0.170 \pm 0.006~R_{\odot}$, and $R_{2} = 0.156 \pm 0.006~R_{\odot}$, respectively. Based on the luminosity ratio from the light curve modeling, the effective temperatures of two components are also estimated. In comparison with the stellar evolution models, the radii and effective temperatures of two components are all below the isochrones. The radius deflation might be mainly biased by a small radial velocity (RV) data or (and) a simple correction on RVs, while the discrepancy in effective temperature might be due to the enhanced magnetic activity in this binary. 
\end{abstract}

\begin{keywords}
binaries: eclipsing -- binaries: spectroscopic -- stars: low-mass -- stars: individual: TMTSJ0803
\end{keywords}



\section{Introduction}
Low-mass stars are the most common stellar objects in our galaxy \citep{2006AJ....132.2360H, 2012ApJ...757...42F} and understanding them is thus clearly an important endeavour. \cite{2006AJ....132.2360H} found that at least $\sim$ 70\% of stars within 10 pc of the Sun are M dwarfs with $M \le 0.6~M_{\odot}$. Such low-mass stars are useful probes of their structures \citep[e.g.][]{2008ApJ...673..864J}, kinematics \citep[e.g.][]{2007AJ....134.2418B}, and chemical evolution \citep[e.g.][]{2012MNRAS.422.1489W, 2022ApJ...927..123S}. The fundamental properties of M dwarfs have become an essential component to studies of the initial mass function \citep[e.g.][]{2023Natur.613..460L}. M dwarfs are also attractive targets for the identification and characterization of exoplanets. Due to smaller sizes, it is easier to find Earth-size planets around M dwarfs compared to FGK stars \citep[e.g.][]{2008PASP..120..317N, 2009Natur.462..891C}. 

In comparison with solar-type stars, stellar theory is not well understood for low-mass stars, especially in the regime of 0.08--0.3 $M_{\odot}$, because of complex and varied physics inside the stars and active magnetic phenomena on  their surfaces\citep{2001ApJ...559..353M}. The measurements of fundamental properties (mass, radius and effective temperature) are crucial for calibrating stellar evolution models. Careful observations of detached double--lined eclipsing binaries can result in model-independent mass and radius estimates to a precision better than 1\% in some cases \citep[e.g.][]{2009ApJ...691.1400M, 2011ApJ...728...48K}. The first precise determinations of the fundamental parameters, obtained by \citet{2002ApJ...567.1140T} and \citet{2003AA...398..239R} for early and mid-M dwarfs in two eclipsing systems YY Gem and CU Cnc, respectively, indicated that their radii are inflated by up to 20\% with respect to the model predictions. The discrepancies between the observed radii of CM Dra and that predicted by stellar models were also noticed \citep[e.g.][]{2009ApJ...691.1400M, 2012ApJ...760L...9T}, however, CM Dra is only inflated by about 2\% compared to the prediction based on an older stellar age and a near-solar metallicity \citep{2014A&A...571A..70F}. The M dwarfs with $M \gtrsim 0.35~M_{\odot}$ with inflated radii is well established by observations \citep{2011ApJ...728...48K, 2013ApJ...776...87S, 2018MNRAS.476.5253C}. In the range $M \lesssim 0.35~M_{\odot}$, however, the mass and radius of fewer stars can be accurately determined due to low brightness of cool and small objects \citep{2014MNRAS.437.2831Z, 2017ApJ...836..124D}. 

Recent studies suggested that most of the late M dwarfs in the fully convection regime follow the theoretical mass--radius relation \citep{2018AJ....155..114H, 2019AA...625A.150V, 2022MNRAS.513.6042M}. Nevertheless, the effective temperatures measured for some low-mass cool stars were reported to be lower than those predicted by models \citep{2005ApJ...631.1120L, 2009ApJ...691.1400M, 2015MNRAS.451.2263Z, 2018AJ....155..114H}. 
The study on white-dwarf and M dwarf binaries by \cite{2018MNRAS.481.1083P} indicates that there is a 5\% systematic bias towards larger radii for a sample of fully and partially convective low-mass stars, while the temperature measurements for the fully convection stars are in agreement with the theoretical predictions. The possible different trends of temperatures and radii between fully and partially convective stars may be caused by important transitions in inner structures of these stars \citep{1997A&A...327.1039C}. This could be the explanation of a discontinuity in the effective temperature--radius relation discovered by \cite{2019MNRAS.484.2674R} for M dwarfs.

It remains unclear whether the radius and temperature discrepancy is attributed to interior structure of low-mass stars or a systematic effect specific to the short-period binary systems. Stellar magnetic activity is a popular explanation of the inflation of radii of low-mass stars \citep{2001ApJ...559..353M, 2007A&A...472L..17C, 2014ApJ...789...53F}. The activity hypothesis is favored by observations of the short period binary systems \citep{2009ApJ...691.1400M, 2011ApJ...728...48K, 2013ApJ...776...87S}. The tidal interaction in such systems can give rise to a fast rotation of the companions, which is expected to generate strong magnetic fields and lead to larger stellar radii. The magnetic field might explain the radius inflation for M dwarfs with larger mass, however, its strength is unlikely stable in the fully convective interior of these stars \citep{2014ApJ...789...53F}. Other possible mechanisms include metallicity effects \citep{2006ApJ...644..475B, 2019AA...625A.150V} and magnetic star-spots on stars \citep[e.g.][]{2010ApJ...718..502M}. A final solution to the discrepancies may be a combination of the above factors which likely all contribute to the interior structure of fully convective stars \citep{2014A&A...571A..70F}.

In order to probe different effects individually and in aggregate, more sample of low-mass eclipsing binary systems are needed. In this paper we report the study of a short-period eclipsing binary TMTS J08032285+3930509 (dubbed as TMTSJ0803) discovered during the TMTS survey \citep{2022MNRAS.509.2362L}, which consists of two detached M dwarfs in the fully convective mass range. As M dwarf system is very useful for testing theoretical models of low-mass stars, we utilize multicolor photometry from different telescopes and radial velocity (RV) measurements to constrain the masses and radii of both components of this binary system. 

The paper is structured as follows. In Section \ref{sec:obs}, we describe the discovery, the follow-up photometric observations, and collection of the spectra. The light curve and period of the binary system is examined in Section \ref{sec:prd}. In Section \ref{sec:ctr}, we estimate the properties of the binary via SED fitting and analysis of the LAMOST spectrum. In section \ref{sec:BM}, we calculate the absolute physical parameters of TMTJ0803 by a joint analysis of the light curves and RV data. In Section \ref{sec:discu}, we discuss the implications of these parameters in regards to the existing theoretical stellar models. We summarize in Section \ref{sec:con}.

\section{Observations}
\label{sec:obs}

\subsection{Photometric data from TMTS, SNOVA, TESS and TNT}
\label{sec:phot} 

TMTS is a photometric survey with four 40-cm optical telescopes located at Xinglong Observatory in China. The survey operates in such an observation mode: uninterrupted observing of the LAMOST areas for the whole night with a cadence of about 1 min, resulting in discoveries of many interesting short-period variables and eclipsing binaries \citep[see more details in][]{2020PASP..132l5001Z, 2022MNRAS.509.2362L, 2023MNRAS.523.2172L, 2024NatAs.tmp...35L, 2024MNRAS.528.6997G}. A primary eclipse of TMTSJ0803 was observed on 2020 Jan. 15 in the first year TMTS survey (see Table \ref{tab:obs}). The system TMTSJ0803, alternative names are LP 208-19 and 2MASS J08032307+3930558, were first identified as an eclipsing binary by \cite{2013AJ....146..101P}. Based on the LAMOST spectrum and the continuous light curve observed on a whole night, we believe that TMTSJ0803 is a double M dwarfs binary system with a short period ($\sim$ 5.32 hours). TMTSJ0803 is the one of the first 12 short period double M dwarfs binary candidates selected from the database of TMTS survey, based on their light curves, (B-R) colors, and photometric and/or spectroscopic temperatures.

Follow-up photometric observations of TMTSJ0803 have been taken by SNOVA and Tsinghua-NAOC 0.8-m telescope \citep[TNT;][]{2008ApJ...675..626W, 2012RAA....12.1585H}. The SNOVA is a 36~cm telescope located at Nanshan Observatory in China, and it is used to monitor TMTSJ0803 in white light (clear band) and standard I band for a total of 8 nights, as shown in Table \ref{tab:obs}. TMTSJ0803 was also monitored in standard R band on 5 nights with the TNT at the Xinglong Observatory in December 2022. To achieve better photometry, we adopted different exposure time, ranging from 60s to 180s, for different telescopes and under different seeing conditions. 

The standard image processing, such as bias correction, flat correction and source extraction are performed with Ccdproc of Astropy \citep{2017zndo...1069648C} and SExtractor \citep{1996A&AS..117..393B}. 
Four comparison stars are used to calibrate the photometry of TMTSJ0803. 

In addition, TMTSJ0803 has also been observed by TESS \citep{2015JATIS...1a4003R} in 3 sectors (20, 47, and 60) with cadences from 30 minutes to 200 seconds (see Table \ref{tab:obs}). For the long cadence observations in the Kepler mission, it was demonstrated by \cite{2017MNRAS.466.2488Z} that the shapes of light curves of short-period (< 1.5 days) eclipsing binaries are influenced by the smearing effect. As the binary has a very short period, long exposure time in TESS observations would cause a significant influence on the shape of light curve of the binary. Therefore, the light curve from sector 60, at a cadence of 200 s, is only used to measure the physical parameters of the binary in Section \ref{sec:BM}. 

\begin{table}
	\centering
	\caption{Observation logs of TMTSJ0803.}
	\label{tab:obs}
	\begin{tabular}{llll} 
		\hline \hline
		Telescope & Instrument & Obs\_Date & Exposure (Bands) \\
		\hline
                 TMTS  & CMOS  &2020-01-15 &60 s (Luminous filter)\\
                 SNOVA  & CCD     &2021-01-18  &60 s (C) \\
                               &              &2021-02-28  &120 s (C) \\ 
                               &              &2021-10-17  &90 s (C)\\
                               &              &2022-01-11  &  90 s (I) \\
                               &              &2022-01-13  & 90 s (I) \\
                               &              &2022-12-16  & 90 s (C) \\
                               &              &2022-12-21  & 90 s (C) \\
                               &              &2022-12-24  &180 s (I)\\
               TNT     & CCD     &2022-12-27  & 60 s (R) \\
                               &              &2022-12-28  & 90 s (R) \\
                               &              &2022-12-29  & 70 s (R) \\
                               &              &2022-12-30  & 70 s (R) \\
                               &              &2022-12-31  & 70 s (R) \\
                 TESS   & CCD   &2019/Sector 20 &1800 s\\
                        &       &2021/Sector 47  &600 s\\
                        &       &2022/Sector 60  &200 s\\
                 LAMOST &Spectrograph  &2014-12-25 &5400 s \\
                 APOGEE &Spectrograph  &2012-12-02 &1502x2 s  \\
                                 &                       &2013-03-30 &2002x2 s  \\
                                 &                       &2016-10-29 &2236x2 s  \\
                                 &                       &2016-10-30 &3131x2 s  \\
                                 &                       &2016-12-05 &4025x2 s  \\
                                 &                       &2016-12-06 &4249x2 s  \\
                                 &                       &2017-02-01 &4025x2 s \\
		\hline
		\multicolumn{4}{l}{\footnotesize C represents the white light without filter in the telescope.}
	\end{tabular}
\end{table}

\subsection{Spectroscopic data from LAMOST}
\label{sec:lamost}

An optical spectrum of TMTSJ0803 was observed on 26th December 2014 by LAMOST \citep{2012RAA....12.1197C} in low resolution mode ($R\sim$ 1800; wavelength range 370 -- 900 nm; \citealt{2015RAA....15.1095L}). This LAMOST spectrum has a bad quality in the blue arm due to the faintness of two these late M dwarfs at the blue end. While the average signal-to-noise (SNR) ratio of the spectrum is $\sim$ 66 in the red end. The H$_{\alpha}$ emission can be clearly seen in Figure \ref{fig:spec-fitting}, where this emission feature observed at an orbital phase of about 0.48 was identified. The presence of prominent H$_{\alpha}$ emission line indicates that TMTSJ0803 could be very active (see more discussions in Sect. \ref{sec:chro}).

Based on the LAMOST spectrum, the atmospheric parameters of TMTSJ0803 were provided in LAMOST DR7\footnote{http://dr7.lamost.org/v1.2/search} as,  effective temperature $T_{\rm{eff}} = 3003\pm156$ K, surface gravity log $\it{g}$ = 5.5 $\pm$ 0.39, metallicity [M/H] = --0.62 $\pm$ 0.52 dex. However, we reexamine the atmospheric parameters of TMTSJ0803 using our own methodology in Sect. \ref{sec:syn}. 

\subsection{Spectroscopic data from APOGEE }
\label{sec:apogee}
TMTSJ0803 was also observed by the APOGEE (Apache Point Observatory Galactic Evolution Experiment) project in high-resolution mode, involving the APOGEE-N (north) spectrograph \citep{2019PASP..131e5001W} which cover a spectral range of 1.51--1.7 $\mu$m with an average resolution of R $\sim$ 22,500 \citep{2010SPIE.7735E..1CW, 2019PASP..131e5001W}. A total of seven infrared spectra can be extracted from the APOGEE DR16 database \citep{2020ApJS..249....3A, 2020AJ....160..120J}. Among them, four spectra were observed in 2016, while other three spectra were taken in 2012, 2013 and 2017, respectively, as listed in Table \ref{tab:obs}. 
The multiple-epoch spectra enable the measurements of RVs for individual components of TMTSJ0803 (see Table \ref{tab:RV}).

As the APOGEE DR16 pipeline constructed cross-correlation function (CCF) for each infrared spectrum, thus an automated code called apogeesb2\footnote{https://github.com/mkounkel/apogeesb2} can be used to measure RV for each component by deconvolving the custom CCF \citep{2021AJ....162..184K}. To identify the primary and secondary components, the apogeesb2 analyzes each CCF by utilizing the autonomous Gaussian deconvolution Python routine, GaussPy \citep{2015AJ....149..138L}. In this work, the log$\alpha$ parameter is set to 3.1 to 3.8, depending on different spectrum rather than the recommended value 1.5 given in \cite{2021AJ....162..184K}. This is due to the relatively low quality of CCFs resulting from the low SNR of the APOGEE spectra. 

The exposure time of each spectrum is usually longer than an hour except for the earliest spectrum observed in 2012. As mentioned in Section \ref{sec:phot}, the phase smearing effect caused by long exposure time can not be ignored. To evaluate the smearing effect on RV, we simply assume an identical sine curve for both primary and secondary stars because they have almost the same stellar mass as discussed in Section \ref{sec:discu}. Comparing the mean RV within the total exposure time with the one at the middle time, we then obtain the corrected RV value as listed in Table \ref{tab:RV}. 

\begin{figure}
	\includegraphics[width=0.48\textwidth]{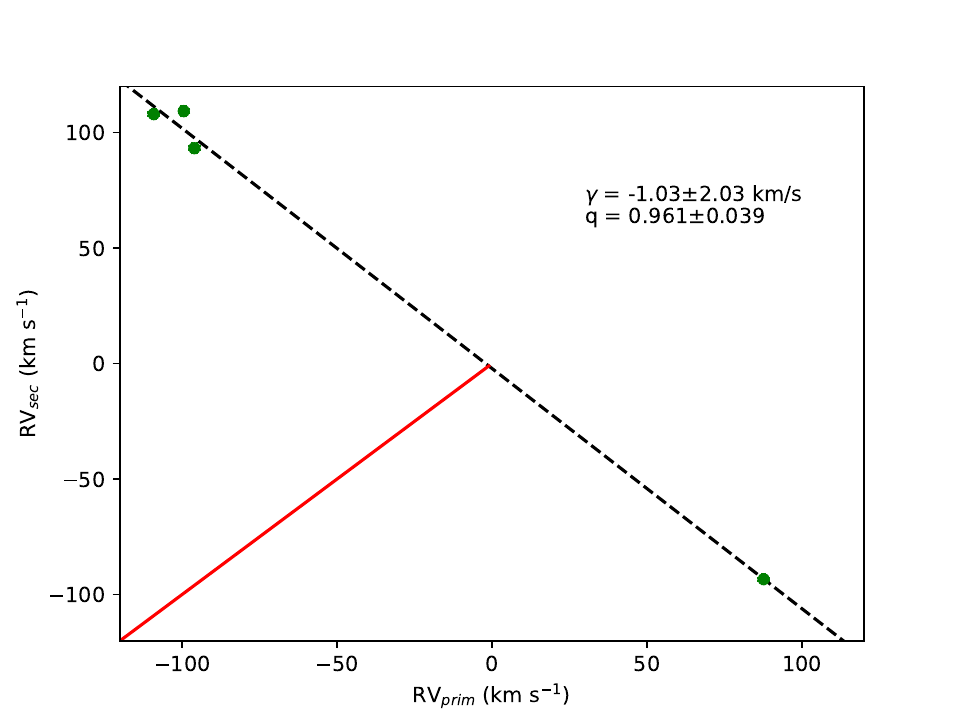}
    \caption{The Wilson plot. Green dots show velocities of the primary relative to those of the secondary. Black dash line shows the best fit to the data; the slope of this line relates to the mass ratio. Red line is the line of equality between $RV_{\rm{prim}}$ and $RV_{\rm{sec}}$; the intersection of these two lines corresponds to the barycentric velocity of the system. }
    \label{fig:wilson}
\end{figure}

Although there is an insufficient number of epochs to construct a full orbital velocity variation, current measurements allow us to determine the mass ratio ( $q = M_{2}/M_{1}$) and the central velocity ($\gamma$) for this binary by constructing a Wilson plot \citep{1941ApJ....93...29W}. We find that the mass ratio and the central velocity are $q=0.961\pm0.039$ and $\gamma=-1.0 \pm 2.0$ km s$^{-1}$, respectively. In Figure \ref{fig:wilson}, the RV pairs from four epochs are used to measure $q$ and $\gamma$. The first RV pair from epoch BJD 2456264.0093 is firstly removed because it is an outlier in the phased RV curves. We find that the absolute RVs (> 57 km s$^{-1}$) at the first epoch are much higher than the expected values ($\sim$ 7 km s$^{-1}$) at phase 0.51. The contradiction might be caused by a wrong measurement of RVs at the first epoch. Two spectra observed at epoch BJD 2456381.7143 and 2457690.9798 are also abandoned, because the correction for phase smearing has a very large effect ($\Delta$RV/RV$_{1,2}$$^{*}$ > 100\%) on their RVs and makes the RVs from these two spectra rather unreliable.
Therefore, in Table \ref{tab:RV} the first three RV pairs are ignored when the RV curves are used to constrain the semi-major axis ($a$) in Section \ref{sec:BM}. 

\begin{table}
	\begin{threeparttable}[t]
	\centering
	\caption{Radial velocities from the APOGEE spectra.}
	\label{tab:RV}
	\begin{tabular}{lrrcrrc} 
		\hline \hline
		BJD & RV$_{1}$ & RV$_{1}$\tnote{*} & $\sigma_{1}$ & RV$_{2}$ & RV$_{2}$\tnote{*} & $\sigma_{2}$\\
		\hline
		--2450000 d &km s$^{-1}$ &km s$^{-1}$ &km s$^{-1}$ &km s$^{-1}$ &km s$^{-1}$ &km s$^{-1}$\\
		\hline
		6264.0093 &--52.55 &--57.60 & 0.66  &58.75  &64.39 & 0.65 \\
		6381.7143 &--38.04 &--14.74 & 3.05  &43.50  &16.85 & 3.22  \\
		7690.9798 &--64.05 &--28.43 & 0.62  &52.07  &23.11 & 0.92  \\
		7691.9455 &--77.19 &--96.00 & 0.81  &74.98  &93.25 & 0.97 \\
		7727.8792 &--75.17 &--109.13 & 0.63 &74.39  &108.00 & 0.72 \\
		7728.8839 &75.02   &87.79   & 0.57  &--79.91  &--93.51 & 0.60 \\
		7785.7522 &--72.40 &--99.36 & 0.72  &79.58  &109.28 & 0.83  \\
		\hline
	\end{tabular}
	\begin{tablenotes}
	\item[*] RV$_{1}$ stands for RV of the primary star, while RV$_{2}$ stands for that of the secondary star. Both RVs are corrected based on the exposure times.
	\end{tablenotes}
	\end{threeparttable}
\end{table}

\section{Period Study}
\label{sec:prd}

\begin{figure*}
	\includegraphics[width=1.0\textwidth]{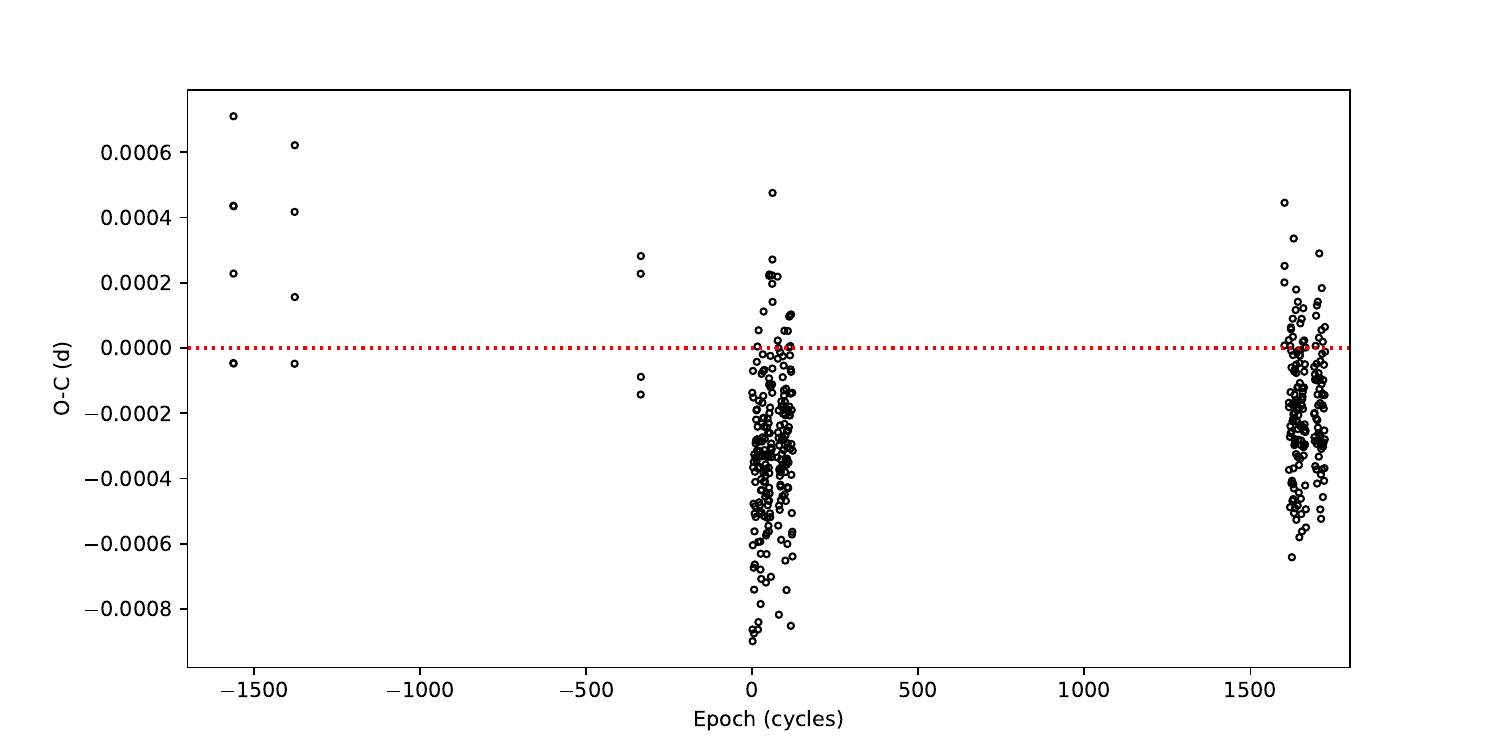}
    \caption{The corresponding O $-$ C residuals after the linear fitting.}
    \label{fig:ephemeris}
\end{figure*}

To derive the orbital ephemeris of TMTSJ0803, we calculate the time of minimum light by fitting its light curve by a Gaussian function. A total of 389 minimum light epochs, 192 primary and 187 secondary ones, are obtained and listed in Table \ref{tab:ephemeris}. With the least-square method, a linear ephemeris is derived by fitting the minimum light time as:
 \begin{equation}
 \begin{split}
 \label{eq:min}
    {\rm{Min.I}} = 2459579.765702(\pm0.000009)~\\
                          +~0.221871501(\pm0.000000007) \times \rm{E}
 \end{split}
 \end{equation}
Figure \ref{fig:ephemeris} shows the linear fitting and the corresponding observed minus computed O $-$ C residuals as a function of the epoch number E. It should be noted that the minimum-light time has a large error (more than 0.0007) and those values derived from eclipses without enough observation data are abandoned. 

\begin{table}
	\centering
	\caption{Minimum times of TMTSJ0803.}
	\label{tab:ephemeris}
	\begin{tabular}{lccc} 
		\hline \hline
		BJD &Error &Epoch  &Filter\\
		--2457000 d  &   &  &\\ 
		\hline
                 2233.20237  &0.00001 &--1562.0 &SNOVA C\\
                 2233.31358  &0.00001 &--1561.5 &SNOVA C\\
                 ... &... &... &...\\
                 2591.08072  &0.00027  &51.0 &SNOVA I\\
                 2591.19197  &0.00050  &51.5 &SNOVA I\\
                  ... &... &... &...\\
		2579.87650  &0.00036 &0.5  &TESS\\
		2579.98671  &0.00030 &1.0 &TESS\\
		2580.09761  &0.00033 &1.5 &TESS\\
		2580.20884  &0.00030 &2.0 &TESS\\
		... &... &... &...\\
                2944.30072  &0.00016 &1643.0 &TNT R \\
                2945.40989  &0.00013 &1648.0 &TNT R \\
        ... &... &... &...\\
        2962.38317  &0.00025  &1724.5 &TESS\\
        2962.49403  &0.00020  &1725.0 &TESS\\
                 
		\hline
		\multicolumn{4}{l}{\footnotesize (This table is available in its entirety in machine-readable form.)}
	\end{tabular}
\end{table}

\section{Characterization of the low-mass eclipsing binary}
\label{sec:ctr}

\subsection{Broad-band photometric characterization}\label{sec:sed}
A total of 19 broad-band photometric datapoints are available for TMTSJ0803, including GALEX\_NUV \citep{2011Ap&SS.335..161B}, SDSS DR12 {\it{u, g, r, i}} \citep{2015ApJS..219...12A}, Pan-STARRS1  {\it{g, r, i, z, y}} \citep{2016arXiv161205560C}, Gaia DR2  {\it{BP, G, RP}} \citep{2018A&A...616A...1G}, 2MASS  {\it{J, H, K$_{\rm{s}}$}} \citep{2006AJ....131.1163S}, WISE W1, W2 \citep{2010AJ....140.1868W}, and TESS \citep{2015JATIS...1a4003R}, which can be used to construct 
the spectral energy distribution (SED) with ARIADNE\footnote{https://github.com/jvines/astroARIADNE} (spectrAl eneRgy dIstribution bAyesian moDel averagiNg fittEr, \citealt{2022MNRAS.513.2719V}), constraining the effective temperature ($T_{\rm{eff}}$), surface gravity (log $\it{g}$), metallicity ([M/H]), distance ($D$) and V-band extinction of the binary system. 
With the Galactic reddening of E(B-V)=0.045 \citep{1998ApJ...500..525S, 2011ApJ...737..103S} and adoption of extinction law from \cite{1999PASP..111...63F}, the best-fit parameters estimated for the system are: $T_{\rm{eff}}$ = 2924 $\pm$ 11 K, log $\it{g}$ =  5.62 $\pm$ 0.11 dex, [M/H] = --0.10 $\pm$ 0.07 dex, $D = 43.57 \pm 0.08$ pc, and A$_{\rm{V}}=0.07 \pm 0.01$ mag. These stellar parameters, except for metallicity, are consistent with the results from a single-star model presented in the next subsection. Here, the error on $T_{\rm{eff}}$ may be underestimated by the methodology. To estimate the possible systematic errors, we calculate the differences between the recommended values from literatures \cite[see Table 3 in][]{2022MNRAS.513.2719V} and $T_{\rm{eff}}$ from the SED-fitting for 13 benchmark M dwarfs (< 3600 K) from \cite{2022MNRAS.513.2719V}, with no systematic offset being found (i.e.,$\sim$ 1 $\pm$ 180 K).

\subsection{Spectroscopic characterization by comparison with synthetic spectra}
\label{sec:syn}
Using the full-spectrum fitting method developed by \cite{2022MNRAS.513.4295K}, we analyze the LAMOST spectrum in binary and single-star spectral models (see details in Section 2.3 in \citealt{2022MNRAS.513.4295K}). For double M-type binary, the synthetic spectra from BT-Settl (AGSS2009) models \citep{2011ASPC..448...91A, 2012RSPTA.370.2765A} rather than the  NLTE MPIA models are used in the analysis. Since there is no reliable absolute flux calibration for the LAMOST spectra, the observed spectrum is flux-renormalized by a fourth-order polynomial recommended from the loss function \citep{2021ApJ...908..131Z}. The synthetic spectrum of binary system is generated as a sum of two Doppler-shifted, normalized, and scaled single-star model spectra which are function of  atmospheric parameters and stellar size. Comparing the synthetic binary spectrum with the observed one yields estimations of optimal $T_{\rm{eff}}$, log $\it{g}$, [M/H], RV of each component, mass ratio $q$ and one set of four coefficients of polynomials. On the other hand, the observed spectrum is also  analysed with a single-star model, which is identical to a binary model when the parameters of both components are equal. 

\begin{figure*}
        \centering
	\includegraphics[width=1.0\textwidth]{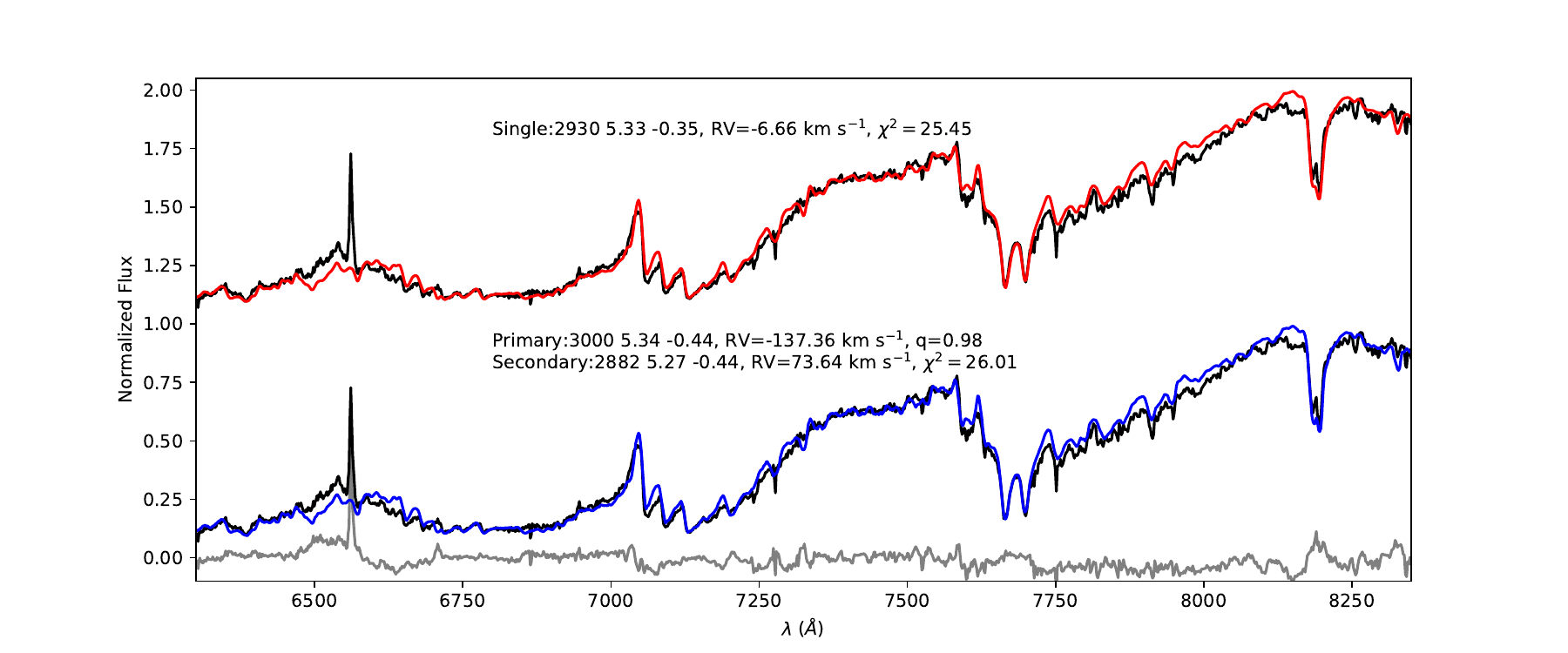}
    \caption{Comparison of the single-star (offset + 1.0) and binary model fits for TMTSJ0803. The difference of observed spectrum and single-star model spectrum is shown as gray line.}
    \label{fig:spec-fitting}
\end{figure*}

Owing to the quality of the spectrum in the blue end, only the spectrum of the red end ranging from 6800 \AA~to 8500 \AA~is fit by models, as shown in Figure \ref{fig:spec-fitting}. As the mass ratio $q$ is measured from the RVs data discussed in Section \ref{sec:apogee}, we fix $q=0.96$ in order to find solution with minimal $\chi^{2}$ in the spectral fitting program. Finally, we obtain the best atmospheric parameters in the single-star model: $T_{\rm{eff}}$ = 2930 K, log $\it{g}$ = 5.33 cgs, and [M/H] = $-$ 0.35 dex. For the binary model, we find $T_{\rm{eff}}$ = 3000 K, log $\it{g}$ = 5.34 cgs, and [M/H] = $-$ 0.44 dex for the primary star, and $T_{\rm{eff}}$ = 2882 K, log $\it{g}$ = 5.27 cgs, and [M/H] = $-$ 0.44 dex for the secondary star. The minimal $\chi^{2}$ of these two fittings are shown in Figure \ref{fig:spec-fitting}. As the $\chi^{2}$ from the binary model is larger then that of the single-star model, therefore, the atmospheric parameters from the single-star model would better constrain the binary properties by fitting to the light curves. Furthermore, by considering uncertainties, our estimated $T_{\rm{eff}}$ is consistent with the effective temperature of 3003$\pm$156 K and 2974$\pm$194 K measured from the LAMOST and APOGEE spectra, respectively.
The effective temperatures of two components in the binary model are consistent with the results (see Table \ref{tab:phy-par}) derived from the light curve modeling (see Section \ref{sec:BM}). 

Errors on $T_{\rm{eff}}$, log $\it{g}$ and [M/H] are provided by the full-spectrum fitting method for both single-star and binary models. However, they are underestimated and nominal, such as 1 -- 10 K in $T_{\rm{eff}}$, and 0.01 dex in log $\it{g}$ and [M/H], because the systematic errors are not included. Since the typical errors have been evaluated by \cite{2022MNRAS.513.4295K} using simulated datasets (see more details in their Sect. 2.3.3), we thus adopt the standard deviation of the test sample as our errors estimation. In this case, for the single-star model and primary component, the typical errors on $T_{\rm{eff}}$, log $\it{g}$ and [M/H] is less than 150 K, 0.1 cgs, and 0.1 dex, respectively. For the secondary component, the typical errors on $T_{\rm{eff}}$, log $\it{g}$ and [M/H] are less than 350 K, 0.2 cgs, and 0.2 dex, respectively.

\section{Binary Modeling}
\label{sec:BM}

\subsection{Modeling light curves and radial velocity shifts}
\label{sec:lcrv}

\begin{figure}
	\includegraphics[width=0.45\textwidth]{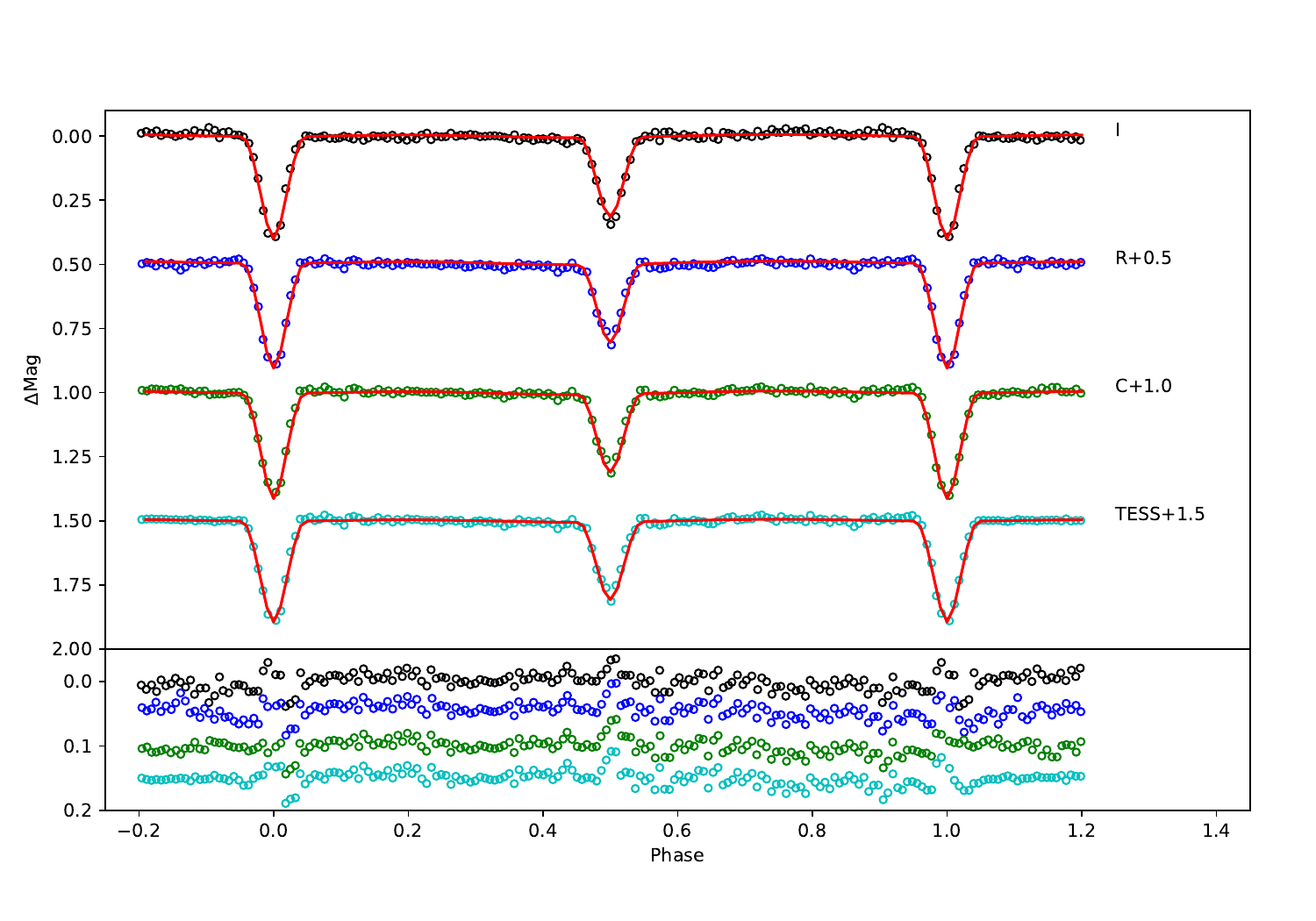}
    \caption{Upper panel: Light curves of TMTSJ0803. The points and solid lines represent the observed and theoretical light curves, respectively. Lower panel: residuals of the best-fit curves  relative to the observed light curves.}
    \label{fig:UIfit}
\end{figure}

To determine the properties of the binary, we use the 2015-version Wilson-Devinney (WD) code \citep{1971ApJ...166..605W, 1979ApJ...234.1054W, 1990ApJ...356..613W, 2012AJ....144...73W, 2014ApJ...780..151W} to fit the light curves and the RVs simultaneously. In the subsequent analysis, the primary and the secondary stars are indicated with the subscripts 1 and 2, respectively. The small asymmetry (O'Connell effect \citep{1951PRCO....2...85O}) seen in the light curves, especially in the I band light curve shown in Figure \ref{fig:UIfit}, could be related to cool spots and/or hot spots on the surface of the stars. Since TMTSJ0803 is a detached system and the Balmer emission lines emerge in the LAMOST spectrum, we thus believe that the asymmetries in the light curves can be attributed to cool spots. Therefore, the binary model with spots (called Model B later) is applied to model the light curves. For comparison, another binary model without spot (called Model A) is also used to fit the light curves. 

In the modelling, the effective temperature of the primary ($T_{\rm{eff,1}} = T_{\rm{m}}$) is fixed to 2930 K according to  the single-star spectral fitting result. According to the early studies by \cite{1967ZA.....65...89L} and \cite{1969AcA....19..245R}, the gravity-darkening exponents $g_{1} = g_{2} = 0.32$ and the bolometric albedo $A_{1} = A_{2} = 0.5$ are adopted in the fitting, respectively. The bolometric and bandpass logarithmic limb-darkening coefficients are interpolated from the tables given in \cite{1993AJ....106.2096V}, \cite{2011A&A...529A..75C} and \cite{2017A&A...600A..30C}. As the mass ratio ({\it{q}}) and the center-of-mass velocity ($\gamma$) were determined from the RV data, we therefore fix them in the binary modeling process. Moreover, the limited RV data can not provide enough constraint on the eccentricity ({\it{e}}), we thus simply assume it to be zero. During the fitting, the adjustable parameters are the semi-major axis of the binary ({\it{a}}), the inclination ({\it{i}}), the effective temperature of the secondary ($T_{\rm{eff,2}}$), the surface potential ($\Omega_{1}$ and $\Omega_{2}$), the phase shift, the dimensionless luminosity of the primary ($L_{1}$), the spot parameters, longitude ($\psi$), spot angular radius ({\it{r}}), and temperature factor ($T_{\rm{s}}/T_{*}$). Since the spot area and the latitude are highly correlated with temperature and radius of the star \citep{2014MNRAS.442.2620Z}, respectively, the latitude of the spot is assumed to be at 90$^{\circ}$($\theta \simeq$ 1.571 in radian). 

\begin{figure}
	\includegraphics[width=0.45\textwidth]{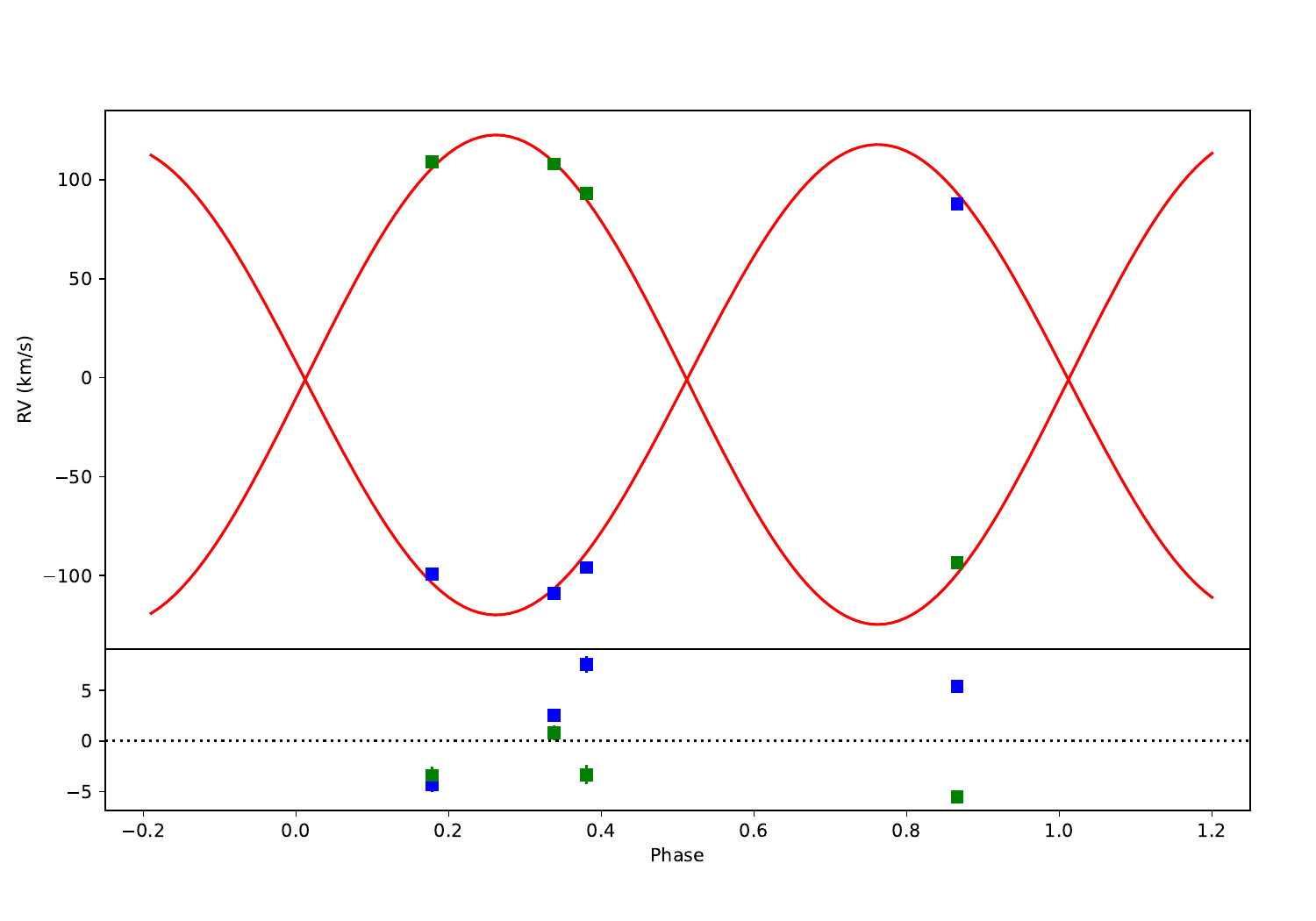}
    \caption{Upper panel: Differential radial velocity curves of the primary (blue squares) and secondary stars (green squares) of TMTSJ0803. The red lines represent corresponding fitting curves. Lower panel: residuals of the best-fit curves relative to the radial velocities.}
    \label{fig:rvfit}
\end{figure}

It is well known that errors on the parameters provided by the WD code are underestimated. To overcome the drawback, firstly the observational errors\footnote{Here the observational error is the standard deviation of the observational points in a small phase bin. The median errors in white light, I band, R band, and TESS data are 0.026, 0.018, 0.029, and 0.020 mag.} are used as weights in our modeling processes to estimate the random errors. To estimate the systematic errors caused by the choices of input physics in the WD model, secondly numerous solutions are generated by choosing different gravity darkening exponents, limb darkening law and coefficients, albedo values, reflection effect, and whether or not to include spots. Here the systematic errors for {\it{i}}, $T_{\rm{eff,2}}$, $\Omega_{1}$, $\Omega_{2}$, {\it{a}}, and four equivalent radius ($r$) for each star are estimated and given in Table \ref{tab:phy-par} (see the second error). We find that the systematic errors is at least twice times larger than the random errors, except for the errors of semi-major axis. Both the random and systematic errors are propagated to the uncertainties of the final absolute parameters, such as mass and radius. Although there is a small effect on some fitted parameter values caused by the spot, two WD models give us the same absolute parameters (see Table \ref{tab:phy-par}). This suggests that the free parameters are well constrained by the multiple light curves in this work.

\subsection{Results}
The best-fit light cures and RV curves (red solid lines) from the binary model and their corresponding $O - C$ residuals are shown in Figure \ref{fig:UIfit} and \ref{fig:rvfit}, respectively. The results of the best-fit models are listed in Table \ref{tab:phy-par}, where the converge solutions without spot are listed in Model A and the solutions with one spot on the secondary are listed in Model B. The smaller value of $\sum (O-C)^{2}_{i}$ in Model B gives us a better solution. This is consistent with the conclusion in Sect. \ref{sec:discu} that the binary system is active from the phenomenon of the Balmer emission lines, X-ray radiation and the flare events. Here, a third body around the binary system is tested, while the contribution of the third light to the total system is estimated to be almost zero in all bands.

Taking the semi-major axis of the binary ({\it{a}}) and the mass ratio ({\it{q}}) into the Kepler's third law ($M_{1}+M_{2}=0.0134a^{3}/P^{2}$) and the equivalent radius ($r = R/a$), we calculate the mass and radius of primary and secondary star, respectively. The units of $M_{1,2}$, $a$, and $P$ in the Kepler's equation should be $M_{\odot}$, $R_{\odot}$, and days, respectively. Although the uncertainties of the final absolute parameters are carefully estimated in the previous section, it is likely underestimated due to the limitation of RV data.

From Table \ref{tab:phy-par}, we find that the ratio of the temperature and luminosity between the secondary and the primary is determined to be $\sim$ 0.96 and 1, respectively. As the low-resolution spectrum observed by LAMOST covers a phase range of 0.14 to 0.84, the light contribution from the secondary star to the observed spectrum can not be neglected. According to the study by \cite{2018PASA...35....8K}, we then calculate the final temperatures: 
\begin{equation}
T_{\rm{eff,1}} =  T_{\rm{m}} + \frac{c\Delta T} {c+1}
\end{equation} 
\begin{equation}
T_{\rm{eff,2}} = T_{\rm{eff,1}} - {\Delta T}
\end{equation}
where ${\Delta T} = T_{\rm{eff,1}} - T_{\rm{eff,2}}$ and $c = L_{2}/L_{1}$ are calculated based on the final solutions from Model B. 
We find that the final $T_{\rm{eff,1}}$ = 2971 K and $T_{\rm{eff,2}}$ = 2869 K are consistent with the temperatures given by the binary spectral fitting method. Both the two empirical relations between effective temperature and spectral type, given by \cite{1991AJ....101..662B} and \cite{2013A&A...556A..15R}, suggest similar spectral types for primary (M5) and secondary star (M5.5). Therefore the detached binary system is composed of two similar late M dwarf (M5 + M5.5) stars. 

\begin{table}
	\begin{threeparttable}
	\centering
	\caption{Light curve solution and physical parameters of the binary TMTSJ0803.}
	\label{tab:phy-par}
	\begin{tabular}{lcc} 
		\hline \hline
		Parameter &Model A (no spot)  &Model B (with spot)\tnote{a}\\
		\hline
        {\it{i}} (deg)       &83.305$\pm$0.04    &83.440$\pm$0.04$\pm$0.12\\
        $T_{\rm{eff,1}}$ (K) &2930 (fixed)\tnote{b} &2930 (fixed)\tnote{b}\\ $T_{\rm{eff,2}}$ (K) &2836$\pm$2         &2825$\pm$2$\pm$150\\
        $\Omega_{1}$         &6.713$\pm$0.056    &7.267$\pm$0.038$\pm$0.12\\
        $\Omega_{2}$         &7.275$\pm$0.039    &7.617$\pm$0.042$\pm$0.19\\
        $r_{1}$ (pole)       &0.158$\pm$0.001    &0.158$\pm$0.001$\pm$0.003\\
        $r_{1}$ (point)      &0.160$\pm$0.001    &0.160$\pm$0.001$\pm$0.003\\
        $r_{1}$ (side)       &0.159$\pm$0.001    &0.159$\pm$0.001$\pm$0.003\\
        $r_{1}$ (back)       &0.160$\pm$0.001    &0.160$\pm$0.001$\pm$0.003\\
        $r_{2}$ (pole)       &0.147$\pm$0.001    &0.145$\pm$0.001$\pm$0.004\\
        $r_{2}$ (point)      &0.148$\pm$0.001    &0.147$\pm$0.001$\pm$0.004\\
        $r_{2}$ (side)       &0.147$\pm$0.001    &0.146$\pm$0.001$\pm$0.004\\
        $r_{2}$ (back)       &0.148$\pm$0.001    &0.147$\pm$0.001$\pm$0.004\\
   $L_{1}/(L_{1}+L_{2})(R)$  &0.597$\pm$0.006    &0.632$\pm$0.005\\
   $L_{1}/(L_{1}+L_{2})(I)$  &0.588$\pm$0.006    &0.624$\pm$0.005\\
   $L_{1}/(L_{1}+L_{2})(C)$  &0.599$\pm$0.006    &0.634$\pm$0.005\\
 $L_{1}/(L_{1}+L_{2})(Tess)$ &0.590$\pm$0.006    &0.626$\pm$0.005\\
		$\theta$ (rad)          &---      &1.571 (fixed)\tnote{b} \\
		$\psi$ (rad)            &---      &0.628$\pm$0.150 \\
		$r$ (rad)               &---      &0.142$\pm$0.012 \\
		$T_{\rm{s}}/T_{*}$      &---      &0.867$\pm$0.039 \\
     $\sum (O-C)^{2}_{i}$           &0.065    &0.058 \\
                 \hline
		Absolute parameters: & &\\
		\hline
		{\it{a}} ($R_{\odot}$)  &1.066$\pm$0.021  &1.066$\pm$0.021$\pm$0.020\\
		$M_{1}$ ($M_{\odot}$)   &0.169$\pm$0.010  &0.169$\pm$0.010 \\
		$M_{2}$ ($M_{\odot}$)   &0.162$\pm$0.016  &0.162$\pm$0.016 \\
		$R_{1}$ ($R_{\odot}$)   &0.170$\pm$0.006  &0.170$\pm$0.006 \\
		$R_{2}$ ($R_{\odot}$)   &0.157$\pm$0.006  &0.156$\pm$0.006 \\
		log $g_{1}$ (cgs)       &5.21$\pm$0.04   &5.21$\pm$0.04  \\
		log $g_{2}$ (cgs)       &5.26$\pm$0.05   &5.26$\pm$0.05  \\
		log $L_{1}$ ($L_{\odot}$)   &-2.695$\pm$0.006  &-2.696$\pm$0.006  \\
		log $L_{2}$ ($L_{\odot}$)   &-2.830$\pm$0.006  &-2.830$\pm$0.006  \\
		\hline
	\end{tabular} 
	\begin{tablenotes}
	\item[a] The systematic errors that are propagated to the uncertainties of the final absolute parameters are estimated for some parameters, such as, {\it{i}}, $T_{\rm{eff,2}}$, {\it{a}} and so on.
	\item[b] The primary temperature and spot latitude are fixed when modeling the light curves and spot on the star surface.
	\end{tablenotes}
	\end{threeparttable}

\end{table}

\section{Discussions}
\label{sec:discu}

\subsection{Activity analysis} \label{sec:chro}
Chromospheric activity and star-spot activity could be triggered by magnetic fields and maintained by a magnetic dynamo. The spectral lines (H$_{\alpha}$, H$_{\beta}$, H$_{\gamma}$, H$_{\delta}$ and Ca II H\&K) are useful diagnostic indicators of chromospheric activity for late-type stars. The chromospheric activity of M stars shows emissions above continuum or core emissions in Balmer lines. 

In our analysis, we first created a subtracted spectrum (the observed spectrum minus the synthetic spectrum from the binary model in Sect. \ref{sec:syn}) based on the observed LAMOST spectrum. When calculating the equivalent width of the H$_{\alpha}$ line (EW), we integrate the emission profile using the following formula 
\begin{equation}
{\rm{EW}} =  \int_{\rm{line}} \frac{F_{\lambda}-F_{c}}{F_{c}} d\lambda
\end{equation} 
where $F_{\lambda}$ and $F_{c}$ represent the fluxes of spectral line and the continuum. Comparing the criteria (0.75 \AA) for determining the chromospheric activity of M-type stars \citep{2011AJ....141...97W}, the EW of H$_{\alpha}$ (2.63 \AA) confirms the eclipsing binary is active. 

The ratio of the H$_{\alpha}$ luminosity to the bolometric luminosity of the star, $L_{\rm{H_{\alpha}}}/L_{\rm{bol}}$, enables a better mass-independent comparison between activity levels in M dwarfs than EW alone \citep{2017ApJ...834...85N}. If we simply consider TMTSJ0803 as a single star, $L_{\rm{H_{\alpha}}}/L_{\rm{bol}}$ could be easily calculated: $L_{\rm{H_{\alpha}}}/L_{\rm{bol}}$ = EW $\times$ $\chi$ by adopting $\chi$ factor from \cite{2014ApJ...795..161D}. Our result ($L_{\rm{H_{\alpha}}}/L_{\rm{bol}} \sim 0.65 \pm 0.13 \times 10^{-4}$) is consistent with the values for most of rapidly rotating and full convective stars \citep{2014ApJ...795..161D, 2017ApJ...834...85N}.

Like the spectral lines, coronal X-ray emission is also a useful diagnostic of stellar activity. A close relation between the surface magnetic flux and X-ray radiance indicates that X-ray emission is a reliable proxy of magnetic activity \citep{2016Natur.535..526W}. Crossmatching with the updated ROSAT point-source catalogue \citep[2RXS][]{2016A&A...588A.103B} within 2 arcsecond, we find star ROSAT 2RXS J080322.0+393050 has a X-ray spectrum in the 0.1-2.4 keV energy band. With the SPIDERS program \citep{2017MNRAS.469.1065D, 2020A&A...636A..97C}, the observed instrumental ROSAT count rates can be converted into physical flux at 6.42965 $\pm$ 2.40437 $\times 10^{-13}$ mW/m$^{2}$. Given a parallax distance of 43.584 $\pm$ 0.10 pc \citep{2021AJ....161..147B, 2021A&A...649A...1G}, the flux can be used to calculate the X-ray luminosity which is $L_{X} = 3.8175 \pm 1.4276 \times 10^{-5} L_{\odot}$. Combining with the Bolometric luminosity from the results of SED fitting in Section \ref{sec:sed}, we finally obtain $L_{X}/L_{\rm{bol}} = 0.0159 \pm 0.0059$. Comparing with the fractional X-ray luminosity ($L_{X}/L_{\rm{bol}} < 0.005$) of rapidly rotating fully convective stars, we find that the luminosity of TMTSJ0803 is 3 times brighter than that of the strongest X-ray stars in the catalog of \cite{2011ApJ...743...48W} and much higher than the mean saturation level of $10^{-3}$. This suggests that stellar magnetic activity of fully convective stars could be significantly enhanced in a very close-by binary system such as TMTSJ0803. It is highly possible that the two components of this binary system have synchronized their rotation periods with the orbital period (i.e., $\sim$ 5.32 hours) via tidal interaction. This is consistent with the discoveries that tidal locking leads to larger magnetic fields due to faster rotation rate \citep{2013ApJ...776...87S, 2022A&A...668A.116G}.

Flares are sudden and violent events that release magnetic energy and hot plasma from the stellar atmosphere. It is an indicator of the inherent activity of M dwarf stars. A clear flare event emerged around phase 0.23 on BJD 2459957.4416 was observed by TESS in sector 60. The flare duration was found to be $\sim$50 min. Two more flares located around phase 0.69 and 0.83 were observed by TESS in sector 20 and 47, respectively. Those flares confirm that the system has stellar activity.

\subsection{Comparisons with other M dwarf systems and stellar evolution models}

\begin{figure*}
	\includegraphics[width=1.0\textwidth]{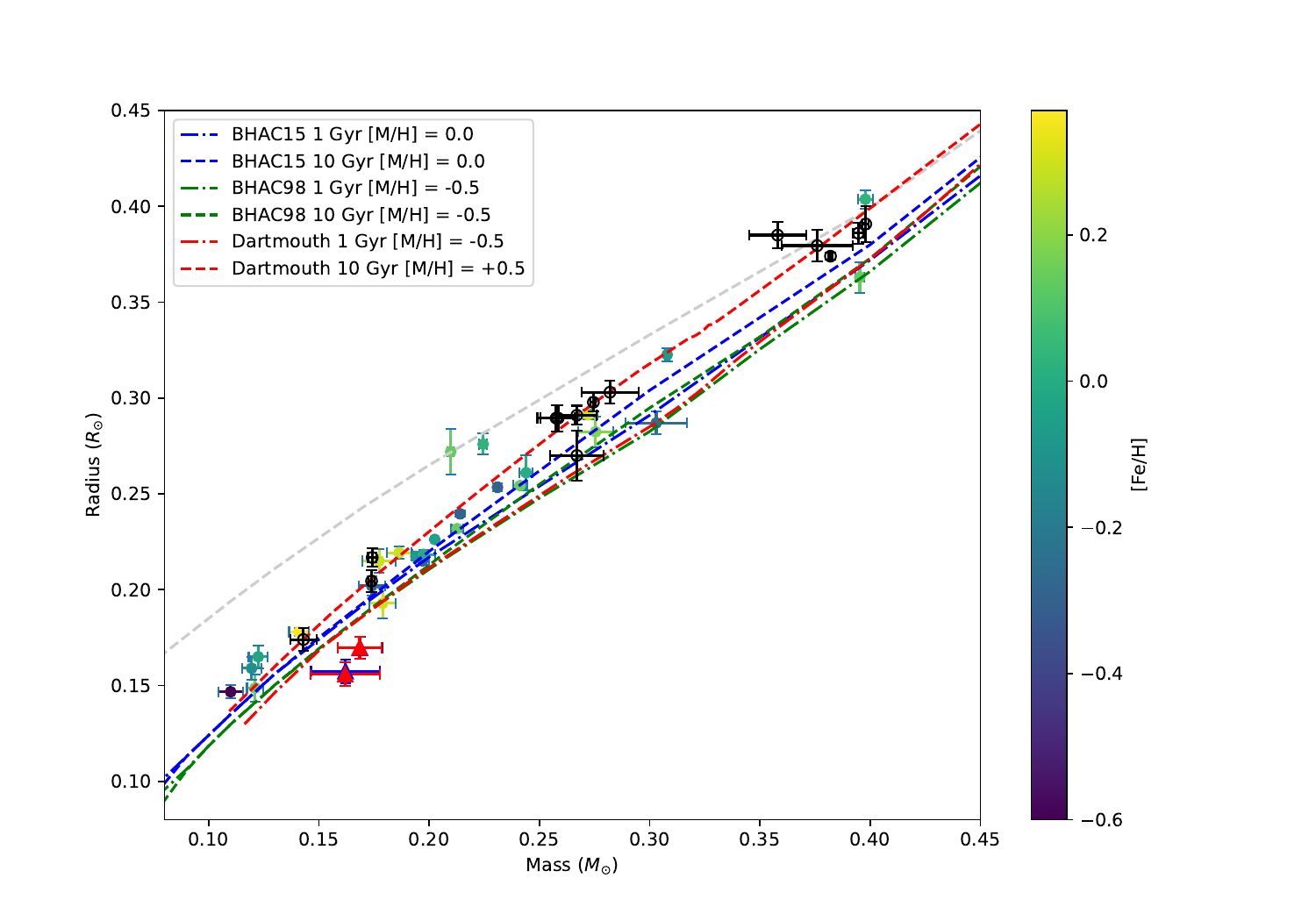}
    \caption{Mass--radius diagram for M dwarfs with 0.09 $M_{\odot} < M < 0.40~M_{\odot}$. The color scale of the points indicates different metallicity. Large filled blue and red triangles show the components of TMTSJ0803 in Model A and B, respectively. Smaller circles show other M dwarfs with parameters given in Table \ref{tab:phy-par}. Black open circles are used for systems without a measured metallicity. The red dash and dot-dash lines show theoretical mass--radius relations from the Dartmouth \citep{2008ApJS..178...89D} models with different ages and metallicities. The blue and green lines represent the solar metallicity isochrones of BHAC15 \citep{2015A&A...577A..42B}, while the blue and green dash lines represent the sub-solar metallicity ([Fe/H] = --0.5 dex) isochrones of BHAC98 \citep{1998A&A...337..403B}  age 1 and 10 Gyr.}
    \label{fig:mr}
\end{figure*}

\begin{figure}
	\includegraphics[width=0.48\textwidth]{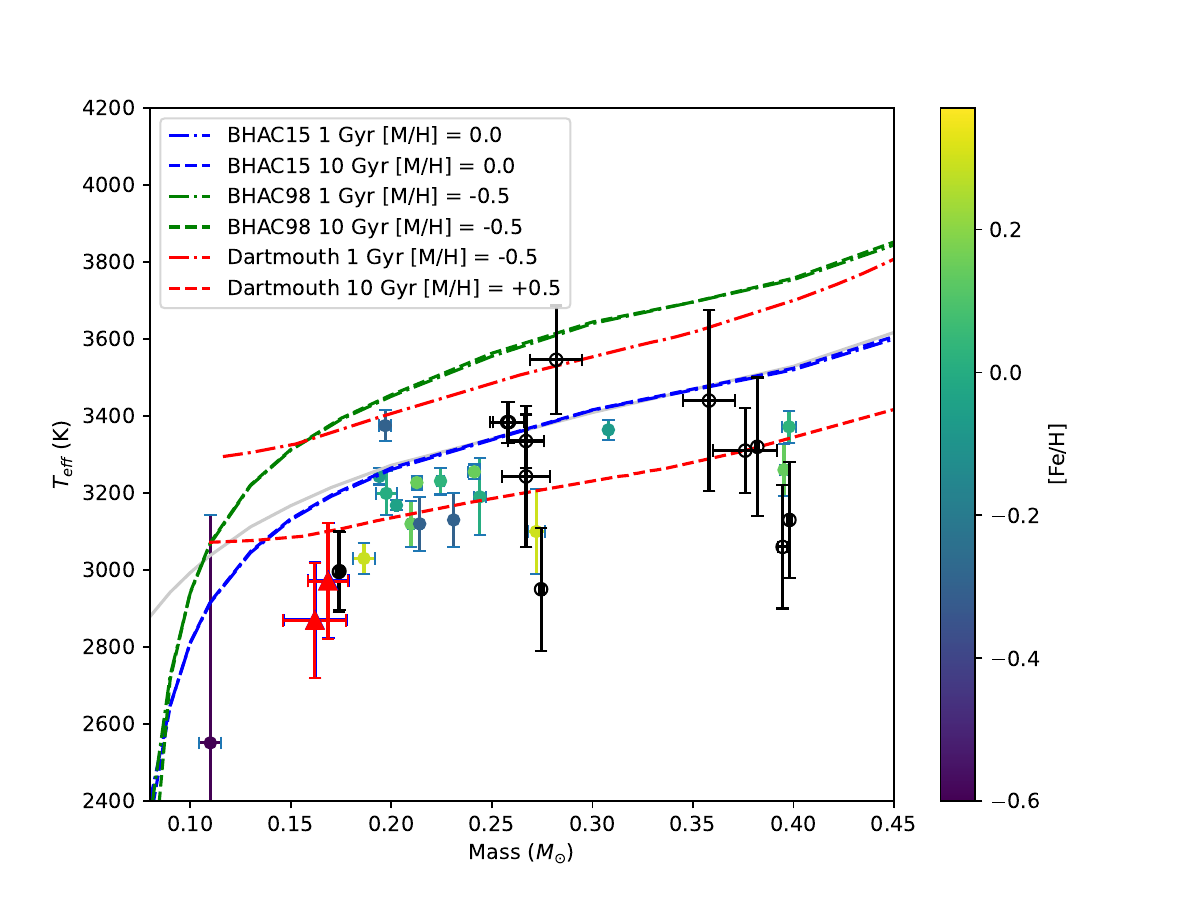}
    \caption{Similar to Figure \ref{fig:mr}, but we show the mass--$T_{\rm{eff}}$ diagram. The observed temperatures are systematically below the theoretical models, with even the 10 Gyr, [Fe/H] = +0.5 dex model being above about the half of the observations.}
    \label{fig:mt}
\end{figure}

\begin{figure}
	\includegraphics[width=0.48\textwidth]{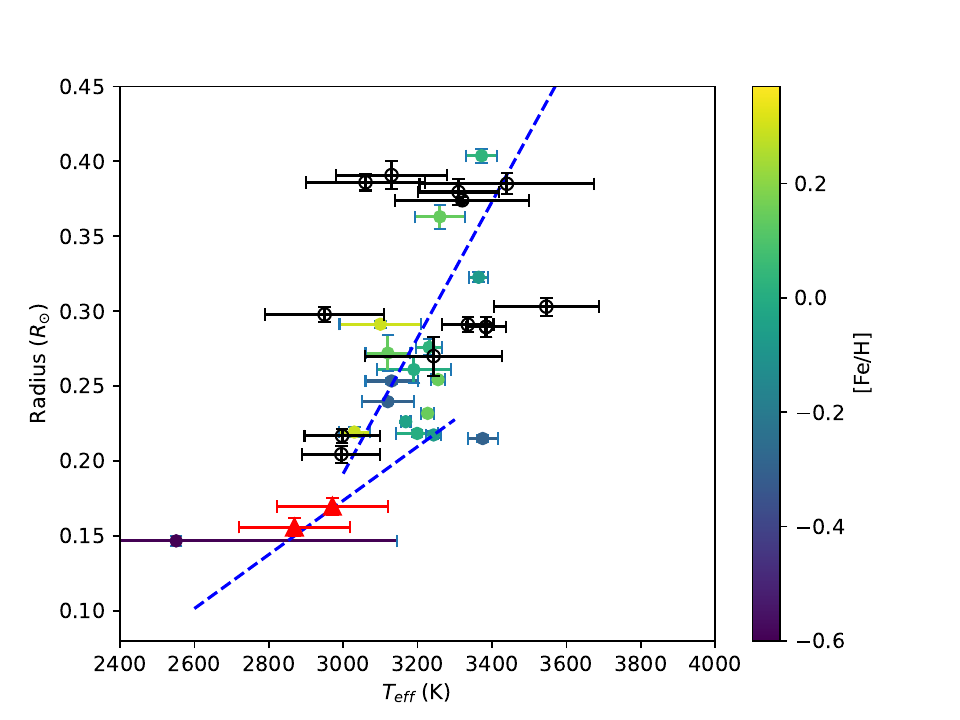}
    \caption{Similar to Figure \ref{fig:mr}, but here we show the $T_{\rm{eff}}$--radius diagram. The dash blue lines represent the discontinuity of the $T_{\rm{eff}}$--radius relation at stellar mass 0.23 M$_{\odot}$ \citep{2019MNRAS.484.2674R}. It looks like that there is no discontinuous behaviour for M dwarfs from binaries.}
    \label{fig:tr}
\end{figure}

Besides TMTSJ0803, there are 41 other M dwarf stars in eclipsing binaries with masses between 0.1 and 0.4 $M_{\odot}$ and with masses and radii measured to have an accuracy better than 5\%. Parameters of these binary systems are collected in Table \ref{tab:ref}. Figure \ref{fig:mr}, \ref{fig:mt}, and \ref{fig:tr} show the mass--radius, mass--$T_{\rm{eff}}$, and $T_{\rm{eff}}$--radius relations for these objects, respectively. Inspecting the mass--radius and mass--$T_{\rm{eff}}$ plots reveals that the isochrones from the Dartmouth theoretical stellar evolution models show the best match with the observations. Although almost all the observed radii are above the age = 1 Gyr isochrone, they are consistent with the old and metal-rich isochrone ([Fe/H] = +0.5 dex, age = 10 Gyr) predicted by the Dartmouth models. This suggests that most of the field stars may be old and/or metal-rich, while stars from open clusters (NGTS J0002-29A, B and PTFEB132.707+19.810A, B) are close to the young isochrone (age = 0.1 Gyr) in the mass--radius plot. Unlike the masses and radii are more consistent with the models, the observed effective temperatures are systematically lower than the models, especially the BHAC98 and BHAC15 models. 

Figure \ref{fig:mr} shows the locations of TMTSJ0803 in Radius$-$Mass relation as inferred from Models A and B, together with those of known sample of double M dwarf binaries (see also Table \ref{tab:phy-par}). Figure \ref{fig:mt} shows distribution of TMTSJ0803 and the other sample in the $T_{\rm{eff}}$--Mass relation. The outstanding feature is that the observed radii of two components of TMTSJ0803 are below the isochrones, while their effective temperatures are not outliers in the M dwarfs sample. Comparison with the stellar evolution models, smaller radii might be due to limited $RV$ data or (and) a simple correction for RVs based only on the exposure times (see Section \ref{sec:apogee}). We find from Table \ref{tab:RV} that the difference between the original RV and corrected RV could be larger than 30 km s$^{-1}$. According to the third Kepler law, the semi-major axis is in proportion to the maximum RV of the binary when we fix the period. In this case, the uncorrected RV may decrease the radius by about 20\% at least. Moreover, there is no RV data in the third quarter of the phase (see Figure \ref{fig:rvfit}). This might also cause bias in the final stellar radius. Therefore, it is not surprising that the radii of two components are slightly below the isochrones.

We find that different methodologies give different metallicities by fitting to the spectrum of TMTSJ0803. According to the parameters given by LAMOST DR7, the metallicity of this binary system is [M/H] = $-$0.62 dex with a large uncertainty. Comparing with the single-star model in Section \ref{sec:syn}, the binary model gives a [M/H] = $-$0.44 dex. From the combined infrared spectrum, we find that the best stellar atmospheric parameters have been provided by the pipeline ASPCAP in APOGEE DR14 \citep{2016AJ....151..144G, 2018ApJS..235...42A} and the system metallicity is [M/H] = $-$0.94 dex. Those suggest that the metallicity of TMTSJ0803 might be more poor than [M/H] = $-$0.35 dex used in this work. In this case, it would not be surprised that the estimated radii and temperatures are all below the isochrones. 

\begin{table*}
	\begin{threeparttable}
	\centering
	\caption{The known M dwarfs in eclipsing binary systems with masses between 0.1 and 0.4 $M_{\odot}$, and with better determinations of masses and radii (i.e., $<$ 5\% error).}
	\label{tab:ref}
	\begin{tabular}{lccccl} 
		\hline \hline
		Star & Mass &Radius &$T_{\rm{eff}}$ & [Fe/H]\tnote{*} & Reference(s) \\
		       & ($M_{\odot}$) & ($R_{\odot}$) & (K) &    & \\
		\hline	
CU Cnc B  &0.3980$\pm$0.0014  &0.3908$\pm$0.0094  &3130$\pm$150  &...  &\citep{2003AA...398..239R} \\
NGTS J0002-29A\tnote{a}  &0.3978$\pm$0.0033  &0.4037$\pm$0.0048  &3372$\pm$41  &+0.03$\pm$0.07  &\citep{2021MNRAS.507.5991S, 2016AA...585A.150N} \\
PTFEB132.707+19.810A\tnote{a}  &0.3953$\pm$0.0020  &0.3630$\pm$0.0080  &3260$\pm$67  &+0.14$\pm$0.04 &\citep{2017ApJ...845...72K} \\	
LSPM J1112+7626A  &0.3946$\pm$0.0023  &0.3860$\pm$0.0055  &3060$\pm$160   &...  &\citep{2011ApJ...742..123I} \\
MG1-2056316B      &0.3820$\pm$0.0010  &0.3740$\pm$0.0020  &3320$\pm$180  &...  &\citep{2011ApJ...728...48K} \\
GJ 3236A          &0.3760$\pm$0.0160  &0.3795$\pm$0.0084  &3310$\pm$110   &...  &\citep{2009ApJ...701.1436I} \\
TYC 3576-2035-1 &0.3580$\pm$0.0130 &0.3850$\pm$0.0070 &3440$\pm$235 &... &\citep{2023MNRAS.521.3405J} \\
LP 661-13A\tnote{b}  &0.30795$\pm$0.00084  &0.3226$\pm$0.0033  &...     &--0.07$\pm$0.10  &\citep{2017ApJ...836..124D} \\
EBLM J1847+39B  &0.3030$\pm$0.0140  &0.2870$\pm$0.0060   &...   &--0.25$\pm$0.06  &\citep{2019AA...626A.119G} \\
TYC 3121-1659-1 &0.2820$\pm$0.0130 &0.3030$\pm$0.0060 &3546$\pm$141 &... &\citep{2023MNRAS.521.3405J}\\
EBLM J1403-32B  &0.2755$\pm$0.0079  &0.2824$\pm$0.0080  &...                  &+0.19$\pm$0.14  &\citep{2019AA...625A.150V} \\ 
LSPM J1112+7626B  &0.2745$\pm$0.0012  &0.2978$\pm$0.0049  &2950$\pm$160   &...  &\citep{2011ApJ...742..123I}	\\
HAT-TR-318-007B\tnote{c}  &0.2721$\pm$0.0042  &0.2913$\pm$0.0024  &3100$\pm$110 &+0.298$\pm$0.08  &\citep{2018AJ....155..114H} \\
TYC 3473-673-1 &0.2670$\pm$0.0090 &0.2910$\pm$0.0050 &3335$\pm$69 &... &\citep{2023MNRAS.521.3405J}\\
HAT-TR-205-003 &0.2670$\pm$0.0120 &0.2700$\pm$0.0130 &3243$\pm$184 &... &\citep{2023MNRAS.521.3405J}\\
1RXS J154727.5+450803B  &0.2585$\pm$0.0080  &0.2895$\pm$0.0068  &...  &...   &\citep{2011AJ....141..166H} \\
1RXS J154727.5+450803A  &0.2576$\pm$0.0085  &0.2895$\pm$0.0068  &...    &...  &\citep{2011AJ....141..166H} \\
HATS551-027A  &0.2440$\pm$0.0030  &0.2610$\pm$0.0090  &3190$\pm$100  &+0.00$\pm$0.20  &\citep{2015MNRAS.451.2263Z} \\
KOI-126B\tnote{d}  &0.2413$\pm$0.0030  &0.2543$\pm$0.0014  &...    &+0.15$\pm$0.08  &\citep{2011Sci...331..562C} \\
CM Dra A\tnote{c}  &0.2310$\pm$0.0009  &0.2534$\pm$0.0019  &3130$\pm$70  &--0.30$\pm$0.12  &\citep{2009ApJ...691.1400M, 2012ApJ...760L...9T} \\
NGTS J0002-29B\tnote{a}  &0.2245$\pm$0.0018  &0.2759$\pm$0.0055  &3231$\pm$35  &+0.03$\pm$0.07  &\citep{2021MNRAS.507.5991S, 2016AA...585A.150N} \\
CM Dra B\tnote{c}  &0.2141$\pm$0.0010  &0.2396$\pm$0.0015  &3120$\pm$70   &--0.30$\pm$0.12  &\citep{2009ApJ...691.1400M, 2012ApJ...760L...9T} \\
KOI-126C\tnote{d}  &0.2127$\pm$0.0026  &0.2318$\pm$0.0013  &...    &+0.15$\pm$0.08  &\citep{2011Sci...331..562C} \\
PTFEB132.707+19.810B\tnote{a}  &0.2098$\pm$0.0014  &0.2720$\pm$0.0120  &3120$\pm$60  &+0.14$\pm$0.04  &\citep{2017ApJ...845...72K} \\	
Kepler-16B\tnote{e}  &0.20255$\pm$0.00066  &0.22623$\pm$0.00059  &... &--0.30$\pm$0.20  &\citep{2011Sci...333.1602D} \\
EBLM J2046+06B  &0.1975$\pm$0.0053  &0.2184$\pm$0.0023  &3199$\pm$57   &+0.00$\pm$0.05  &\citep{2021MNRAS.506..306S} \\
EBLM J0113+31B  &0.1970$\pm$0.0030  &0.2150$\pm$0.0020  &3375$\pm$40  &--0.30$\pm$0.10  &\citep{2022MNRAS.513.6042M} \\
LP 661-13B\tnote{b}  &0.19400$\pm$0.00034  &0.2174$\pm$0.0023  &...     &--0.07$\pm$0.10  &\citep{2017ApJ...836..124D} \\
EBLM J1934-42B  &0.1864$\pm$0.0055  &0.2193$\pm$0.0031  &3030$\pm$41   &+0.29$\pm$0.05  &\citep{2021MNRAS.506..306S} \\
EBLM J1115-36B  &0.1789$\pm$0.0061  &0.1929$\pm$0.0080  &...  &+0.30$\pm$0.14   &\citep{2019AA...625A.150V} \\
EBLM J1013+01B  &0.1773$\pm$0.0077  &0.2150$\pm$0.0060  &...  &+0.29$\pm$0.14   &\citep{2019AA...625A.150V} \\
NGTS J0522-2507B  &0.1742$\pm$0.0019  &0.2168$\pm$0.0048  &2997$\pm$101    &...  &\citep{2018MNRAS.481.1897C} \\
EBLM J2349-32B  &0.1740$\pm$0.0060  &0.2020$\pm$0.0050  &...    &--0.28$\pm$0.06  &\citep{2019AA...626A.119G} \\
NGTS J0522-2507A  &0.1739$\pm$0.0015  &0.2045$\pm$0.0058  &2995$\pm$105    &...  &\citep{2018MNRAS.481.1897C} \\
WTS 19g-4-02069B  &0.1430$\pm$0.0060  &0.1740$\pm$0.0060  &...  &...  &\citep{2013MNRAS.431.3240N}  \\
KIC 1571511B  &0.1410$\pm$0.0045  &0.1779$\pm$0.0020  &...  &+0.37$\pm$0.08  &\citep{2012MNRAS.423L...1O} \\
TYC 2755-36-1 &0.1390$\pm$0.0070 &0.1720$\pm$0.0050 &... &... &\citep{2023MNRAS.521.3405J}\\
GJ 65A  &0.1225$\pm$0.0043  &0.1650$\pm$0.0060  &...  &--0.03$\pm$0.20  &\citep{2016AA...593A.127K} \\
EBLM J1431-11B  &0.1211$\pm$0.0037  &0.1487$\pm$0.0070  &...  &+0.15$\pm$0.14   &\citep{2019AA...625A.150V}\\
GJ 65B  &0.1195$\pm$0.0043  &0.1590$\pm$0.0060  &...  &--0.12$\pm$0.20  &\citep{2016AA...593A.127K} \\
HATS550-016B  &0.1100$\pm$0.0055  &0.1467$\pm$0.0035  &2551$\pm$593  &--0.60$\pm$0.06   &\citep{2014MNRAS.437.2831Z, 2023MNRAS.521.3405J} \\
		\hline
	\end{tabular} 
	\begin{tablenotes}
	\item[*] Normally, the listed metallicity is the value determined spectroscopically for the primary.
	\item[a] NGTS J0002-29 and PTFEB132.707+19.810 are the member of the Blanco 1 and Praesepe open cluster, respectively, and the adopted metallicity is the value for the cluster. 
	\item[b] The metallicity of the LP 661-13 eclipsing binary system was not determined spectroscopically, but was estimated using the absolute $K_{\rm{s}}$ magnitude and the {\it{MEarth}}--$K_{\rm{S}}$ broadband color following \cite{2016ApJ...818..153D}.
	\item[c] Assuming that both components have the same metallicity, the listed value is the binary system metallicities.
	\item[d] KOI-126B and KOI-126C are components of a triply eclipsing hierarchical triple system. The listed metallicity is the value determined spectroscopically for the primary. 
	\item[e] The listed [Fe/H] is the [M/H] value determined spectroscopically for the primary.
	\end{tablenotes}
	\end{threeparttable}
\end{table*}

Unlike a sharp transition identified by \cite{2019MNRAS.484.2674R} for single M dwarfs, the M dwarfs in eclipsing binaries display a roughly linear feature between between stellar radius and $T_{\rm{eff}}$, as seen in Figure \ref{fig:tr}.
The above discrepancy can be explained by several reasons. Firstly, estimating the effective temperatures of a binary system is more challenging in comparison with single star, and the temperatures might suffer a larger bias and uncertainty, such as the system CU Cnc, LSPM J1112+7626, and MG1-2056316 listed in Table \ref{tab:ref}. Secondarily, the data do not show a 'discontinuity' due to the lack of low-mass M dwarfs (< 0.2 $M_{\odot}$) with measurements of $T_{\rm{eff}}$. Finally, there is a possibility that M dwarfs in different environments might have different characteristics. For instance, the stellar activity of individual components could be enhanced in close-binary as discussed in Section \ref{sec:chro}. The strong magnetic fields could inhibit convections in the atmosphere of stars \citep{2012MNRAS.421.3084M}. 

\section{Conclusions} \label{sec:con}
We present a photometric and spectroscopic analysis of the short-period ($\sim$5.32 hours) eclipsing binary TMTSJ0803 detected by the TMTS. The analysis reveals that it is a late M dwarf binary whose components are below the fully convective boundary. Comparing with a normal eclipsing binary, the binary model (Model B) with one spot on the secondary provides the best fit to the light curves and RV data. Under the assumption of the Kepler's third law, we measure the masses and radii of both stars to be $M_{1} = 0.169 \pm 0.010~M_{\odot}$, $M_{2} = 0.162 \pm 0.016~M_{\odot}$, $R_{1} = 0.170 \pm 0.006~R_{\odot}$, and $R_{2} = 0.156 \pm 0.006~R_{\odot}$, respectively. Based on the luminosity ratio from the light curve modeling, the effective temperatures of two components of binary system are determined as $T_{\rm{eff, 1}} = 2971$ K and $T_{\rm{eff, 2}} = 2869$ K, respectively. These are consistent with the results derived by fitting to the LAMOST spectrum with binary model. 

The significant Balmer emission lines seen in the LAMOST spectrum of TMTSJ0803 suggest that this eclipsing binary is very active. Furthermore, we find that TMTSJ0803 has coronal X-ray emission and the fractional X-ray luminosity is $L_{X}/L_{\rm{bol}} = 0.0159 \pm 0.0059$, which is much brighter than that of the typical rapidly rotating fully convective stars. This indicates that the stellar magnetic activity of the fully convective stars could be enhanced in close-by binary environment.

We find that both the radii and temperatures of the two components of TMTSJ0803 are below the isochrones. In comparison with the stellar evolution models, the radius deflation might be mainly biased by the limited RV data or (and) a simple correction for RVs. The effective temperature suppression might be due to enhanced magnetic activity in the binary. To better understand the origin of discrepancy in effective temperature and radius for this system, the higher-precision measurements of RV and photomteric data are required. Combined with high-cadence photometric data, the LAMOST medium resolution survey gives us more opportunities to explore the nature of low-mass eclipsing binaries like TMTSJ0803 from TMTS.

\section*{Acknowledgements}

This work is supported by the National Natural Science Foundation of China (NSFC NO. 12033003, 12288102, and 12203006), MaHuateng Foundation, and Beijing Natural Science Foundation (NO. 1242016). This work was partially supported by Science Program (BS202002), Talents Program (24CE-YS-08), the Innovation Project (23CB061 and 23CB059) and the Popular Science Project (24CD012) of Beijing Academy of Science and Technology, and the Tencent Xplorer Prize. LAMOST is funded by the National Development and Reform Commission and operated and managed by the National Astronomical Observatories, Chinese Academy of Sciences. 

This work has made use of data from the publicly available SDSS data release. Funding for the Sloan Digital Sky Survey IV has been provided by the Alfred P. Sloan Foundation, the U.S. Department of Energy Office of Science, and the Participating Institutions. SDSS-IV acknowledges support and resources from the Center for High Performance Computing  at the University of Utah. The SDSS website is www.sdss4.org.

SDSS-IV is managed by the Astrophysical Research Consortium for the Participating Institutions of the SDSS Collaboration including the Brazilian Participation Group, the Carnegie Institution for Science, Carnegie Mellon University, Center for Astrophysics | Harvard \& Smithsonian, the Chilean Participation Group, the French Participation Group, Instituto de Astrof\'isica de Canarias, The Johns Hopkins 
University, Kavli Institute for the Physics and Mathematics of the Universe (IPMU) / University of Tokyo, the Korean Participation Group, 
Lawrence Berkeley National Laboratory, Leibniz Institut f\"ur Astrophysik Potsdam (AIP),  Max-Planck-Institut f\"ur Astronomie (MPIA Heidelberg), Max-Planck-Institut f\"ur Astrophysik (MPA Garching), Max-Planck-Institut f\"ur Extraterrestrische Physik (MPE), National Astronomical Observatories of China, New Mexico State University, New York University, University of Notre Dame, Observat\'ario 
Nacional / MCTI, The Ohio State University, Pennsylvania State University, Shanghai Astronomical Observatory, United Kingdom Participation Group, Universidad Nacional Aut\'onoma de M\'exico, University of Arizona, University of Colorado Boulder, University of Oxford, University of Portsmouth, University of Utah, University of Virginia, University of Washington, University of Wisconsin, Vanderbilt University, and Yale University.

\section*{Data Availability}

The minimum times of TMTSJ0803 are available in its online supplementary material. The TESS data presented in this paper were obtained from the Mikulski Archive for Space Telescopes (MAST) at the Space Telescope Science Institute (STScI) (https://mast.stsci.edu). Funding for the TESS mission is provided by the NASA Explorer Program directorate.



\bibliographystyle{mnras}
\bibliography{references} 




\appendix




\bsp	
\label{lastpage}
\end{document}